\documentclass[11pt,twoside]{article}
\usepackage{asp2010}

\resetcounters

\begin{document}

\aspvoltitle{To be published in proceedings for Cool Stars 16}
\aspcpryear{in final form by}
\aspvolume{~}
\aspvolauthor{Brown et al.}

\title{Global-scale Magnetism (and Cycles) in Dynamo Simulations of
  Stellar Convection Zones}
\author{Benjamin P.\ Brown$^{1,2}$, 
 Matthew K.\ Browning$^3$, 
 Allan Sacha Brun$^4$, 
 Mark S.\ Miesch$^5$ and
 Juri Toomre$^6$}
 
\affil{$^1$Dept.\ Astronomy, University of Wisconsin, Madison, WI  53706-1582}
\affil{$^2$Center for Magnetic Self Organization in Laboratory and
  Astrophysical Plasmas, University of Wisconsin, Madison, WI 537066-1582}
\affil{$^3$Canadian Institute for Theoretical Astrophysics, University of Toronto, Toronto, ON M5S3H8 Canada}
\affil{$^4$DSM/IRFU/SAp, CEA-Saclay and UMR AIM, CEA-CNRS-Universit\'e Paris 7, 91191 Gif-sur-Yvette, France}
\affil{$^5$High Altitude Observatory, NCAR, Boulder, CO 80307-3000}
\affil{$^6$JILA and Dept.\ Astrophysical \& Planetary Sciences, University of Colorado, Boulder, CO 80309-0440}

\begin{abstract}
  Young solar-type stars rotate rapidly and are very magnetically
  active.  The magnetic fields at their surfaces likely originate in
  their convective envelopes where convection and rotation can drive
  strong dynamo action.  Here we explore simulations of global-scale
  stellar convection in rapidly rotating suns using the 3-D MHD
  anelastic spherical harmonic (ASH) code.   The magnetic fields built
  in these dynamos are organized on global-scales into wreath-like
  structures that span the convection zone.  We explore one case
  rotates five times faster than the Sun in detail.  This dynamo
  simulation, called case~D5, has repeated quasi-cyclic reversals of
  global-scale polarity.  We compare this case D5 to the broader
  family of simulations we have been able to explore and discuss how
  future simulations and observations can advance our understanding of
  stellar dynamos and magnetism.
\end{abstract}

\section{Introduction}
Magnetism is a ubiquitous feature of stars like our Sun.  
The magnetism we see at the surface probably has its origin in stellar
dynamo action arising in the convective envelopes beneath the
photosphere.  There, turbulent plasma motions couple with rotation to
build organized fields on global-scales.  These processes occur in the
Sun as well and are probably the source of the 11-year activity
cycle. Despite intense study, solar and stellar dynamos are poorly
understood, and at present we are unable to reliably predict even
large-scale features of the solar cycle. 

Observations of young, rapidly rotating stars indicate that they have
strong magnetic fields at their surfaces.   There are clearly observed
correlations between rotation and activity which appear to hold
generally for stars on the lower main sequence
\citep[e.g.,][]{Pizzolato_et_al_2003}.  Many of these stars show
cycles of activity as well, though here the dependence on rotation
rate, stellar mass and other fundamental parameters is less clear
\citep[e.g.,][]{Saar&Brandenburg_1999, Olah_et_al_2009}.   
At present even from a theoretical perspective we do not understand
how the stellar dynamo process depends in detail on rotation.

Motivated by this rich observational landscape, we have explored the
effects of more rapid rotation on 3-D convection and dynamo action in
simulations of stellar convection zones.  These simulations have been
conducted using the anelastic spherical harmonic (ASH) code to study
global-scale magnetohydrodynamic convection and dynamo action in
stellar convection zones \citep[e.g.,][]{Clune_et_al_1999,
  Miesch_et_al_2000, Brun_et_al_2004}.  In the past, global-scale
convective dynamo simulations have focused primarily on the Sun, but
now explorations are beginning for a variety of stars, ranging from A-type
\citep[e.g.,][]{Brun_et_al_2005, Featherstone_et_al_2009} to the
M-type dwarfs \citep{Browning_2008}.  

Here we will discuss simulations of G-type stars that rotate more
rapidly than the Sun.  We began these explorations by exploring
convection in hydrodynamic simulations at a variety of rotation rates 
\citep{Brown_et_al_2008}. These simulations capture the convection
zone only, spanning from $0.72\:R_\odot$ to $0.97\:R_\odot$, and take
solar values for luminosity and stratification but the rotation rate
is more rapid. The total density contrast across such shells is about 25.
In those simulations we found that the differential rotation generally
becomes stronger as the rotation rate increases, while the meridional
circulations appear to become weaker and multi-celled in both radius
and latitude.   

These rapidly rotating stars have vigorous dynamos, and 
the magnetic fields created in the dynamos are often organized on
global-scales into banded wreath-like structures
\citep{Brown_et_al_2010a}.   Surprisingly, this
organization occurs in the middle of the convection zone itself,
rather than in a tachocline of penetration and shear between the
convection zone and stable radiative zone beneath.  Many of the
wreath-building undergo quasi-cyclic reversals of magnetic polarity.
Here we explore one of these cyclic dynamos (\S\ref{sec:case D5}),
before putting it in context with other such dynamos
(\S\ref{sec:parameter space}).

\section{Wreaths and Cycles in a Stellar Convection Zone}
\label{sec:case D5}

\begin{figure}[!p]
\begin{center}
\plotone{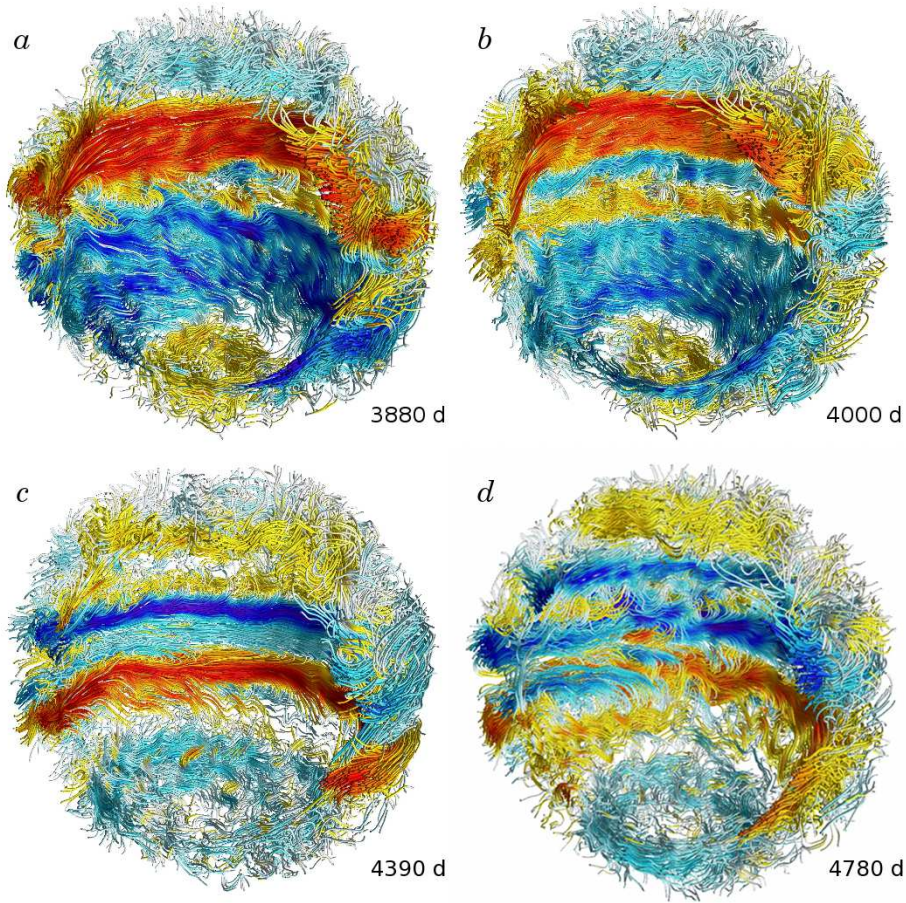}
\caption{Tracing of fieldlines in magnetic wreaths of case
  D5 during a magnetic reversal; volume shown spans slightly more than a full
  hemisphere.
  $(a)$ Shortly before a reversal, with a positive polarity wreath
  above the equator (red tones) and negative polarity below (blue tones). Relic wreaths from
  the previous cycle remain visible in the polar caps. $(b)$~During a reversal,
  new wreaths with opposite polarity form at the equator. 
  $(c)$ When the reversal completes, the polarity of the
  wreaths have flipped, with negative polarity wreath above the
  equator and positive below.  The old wreaths propagate towards the
  poles where they slowly dissipate.
  $(d)$ Mid-cycle, complex non-axisymmetric states are sometimes
  attained but do not always trigger reversals.
  Times of snapshots are labeled, and color tables range from 
  $\pm 25$kG, with peak fields reaching $\pm 40$kG.}
  \label{fig:case D5}
\end{center}
\end{figure}

Our main focus here is on a convective dynamo in a star rotating five
times faster than our Sun currently does, which we call case~D5
\citep{Brown_et_al_2010b}.  Vigorous convection in this simulation
drives a strong differential rotation, which in turns fuels a strong
dynamo.  The magnetic fields created in this dynamo are organized on
global-scales into banded wreath-like structures, as shown
in Figure~\ref{fig:case D5}$a$.  Two wreaths are visible near the
equator, spanning the depth of the convection zone and 
latitudes from roughly $\pm30^\circ$.  The longitudinal field $B_\phi$
dominates the magnetic structures, and the two wreaths have
opposite polarities (positive in northern hemisphere, negative in the
southern).  Magnetic fields meander in and out of each wreath,
connecting them to one another across the equator where small knots of
alternating polarity are visible throughout.  The wreaths are also
connected to high latitudes, where magnetic structures of opposite
polarity are visible; these polar structures are relic wreaths from
the previous global-scale reversal.

We follow one such reversal in Figure~\ref{fig:case D5}.
During the reversal (Fig.~\ref{fig:case D5}$b$), new wreaths of
opposite polarity form near the equator and begin to grow in strength.
After a reversal (Fig.~\ref{fig:case D5}$c$) the new magnetic wreaths
dominate the equatorial region, while the old wreaths
propagate towards the poles.  The origin of this poleward propagation
appears to be a combination of a nonlinear dynamo wave, arising from
systematic spatial offsets between the generation terms for mean
poloidal and toroidal magnetic field, and possibly a poleward-slip
instability arising from magnetic stresses within the wreaths.  Life
near the equator can be quite complex, and at times during the middle
of the cycle states with substantial non-axisymmetry are realized
(Fig.~\ref{fig:case D5}$d$).  In the polar regions, convection begins
to unravel the wreaths from the previous cycle, reconnecting them with
the pre-existing flux there.

\section{Wreath-building Dynamos}
\label{sec:parameter space}

Case D5 is part of a much larger family of simulations that we have conducted
exploring convection and dynamo action in younger suns.  The properties
of this broad family are summarized in Figure~\ref{fig:parameter
  space}$a$.  Indicated here are 26 simulations at rotation rates ranging
from $0.5\:\Omega_\odot$ to $15\:\Omega_\odot$.  At individual rotation
rates (e.g., $3\:\Omega_\odot$), further simulations explore the
effects of lower magnetic diffusivity $\eta$ and hence higher magnetic
Reynolds numbers.  Some of these follow a path where the magnetic
Prandtl number Pm is fixed at 0.5 (triangles) while others sample up
to Pm=4 (diamonds).  The most turbulent simulations have fluctuating
magnetic Reynolds numbers of about 500 at mid-convection zone.
Wreath-building dynamos are achieved in most simulations (17), though
a smaller number do not successfully regenerate their mean poloidal
fields (9, indicated with crosses).  Very approximate regimes of
dynamo behavior are indicated, based on the time variations shown by
the different classes of dynamos.

Detailed studies of cases D3 and D5 indicate that the magnetic wreaths
are built by both the global-scale differential rotation 
and by the turbulent emf arising from correlations in the convection
\citep{Brown_et_al_2010a, Brown_et_al_2010b}.
Generally, the mean longitudinal magnetic field $\langle B_\phi
\rangle$ in the wreaths is generated by the $\Omega$-effect: the
stretching of mean poloidal field by the shear of differential
rotation into mean toroidal field.  Production of $\langle B_\phi
\rangle$ by the differential rotation is typically balanced by
turbulent shear and advection, and by ohmic diffusion on the largest scales.

The mean poloidal field in these simulations is generated by the turbulent emf 
$E_\mathrm{FI} = \langle \vec{u'} \times \vec{B'} \rangle$, where the
fluctuating velocity is 
$\vec{u'}=\vec{u}-\langle \vec{u}\rangle$ and the fluctuating magnetic
fields are $\vec{B'}=\vec{B}-\langle \vec{B}\rangle$.  In cases D3 and
D5, $E_\mathrm{FI}$ is generally strongest at the poleward 
edge of the wreaths, centered at approximately $\pm 20^\circ$ latitude,
whereas the $\Omega$-effect and $\langle B_\phi \rangle$ peak at
roughly $\pm 15^\circ$ latitude.  This spatial offset 
between $E_\mathrm{FI}$ and $\langle B_\phi \rangle$ means that the
turbulent emf is not generally well represented by a simple
$\alpha$-effect description, e.g.,
\begin{equation}
  E_\mathrm{FI} = \langle \vec{u'} \times \vec{B'} \rangle|_\phi \neq \alpha
  \langle B_\phi \rangle
\end{equation}
when $\alpha$ is a scalar quantity.  This is true even when $\alpha$
is estimated from the kinetic and magnetic helicities
present in the simulation.  More sophisticated mean-field models may
do much better at matching the observed emf $E_\mathrm{FI}$, and other
terms in the mean-field expansion may play a significant role; in
particular, the gradient of $\langle B_\phi \rangle$ is large on the
poleward edges of the wreaths where $E_\mathrm{FI}$ is significant.  
During reversals in case D5, both $E_\mathrm{FI}$ and the production of $\langle
B_\phi \rangle$ associated with the $\Omega$-effect surf on the
poleward edge of the wreaths as those structures move poleward.  
This systematic phase shift appears to contribute to that propagation.

\begin{figure}[!t]
\begin{center}
\plotone{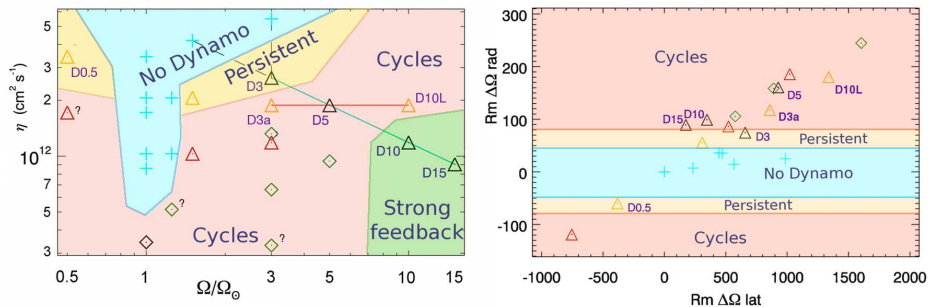}
\caption{Parameter space explored by wreath-building dynamos.  $(a)$
Primary control parameters magnetic diffusivity $\eta$ and rotation
rate $\Omega$ are shown for dynamo simulations at rotation rates
ranging sampling $0.5$--$15\:\Omega_\odot$, with very approximate
dynamo regimes shown and with some cases labeled.  In some regions,
magnetic Reynolds numbers are too low to sustain dynamo action, while
in other regions persistent magnetic wreaths form which do not show
evidence for cycles.  At higher magnetic Reynolds numbers (occurring
here at low $\eta$ or high $\Omega$), wreaths typically undergo
quasi-cyclic reversals.  At the highest rotation rates the Lorentz
force can substantially modify the differential rotation, but dynamo
action is still achieved.  Cases marked with question marks show
significant time-variation but have not been evolved for long enough
to establish cyclic behavior. $(b)$~Plot of magnetic Reynolds numbers
associated with global-scale differential rotation.  The dynamos are
largely driven by differential rotation, and the radial shear
(vertical axis) appears to discriminate between the different dynamo 
regimes.
}
 \label{fig:parameter space}
\end{center}
\end{figure}

As the differential rotation plays a crucial role in these dynamos, we
define magnetic Reynolds numbers associated with the latitudinal shear
at mid-convection zone and the radial shear across the convection
zone:
\begin{equation}
  \mathrm{Rm}~\Delta\Omega~\mathrm{lat} = \frac{\Delta \Omega_\mathrm{lat} R D}{\eta}
\end{equation}
\begin{equation}
  \mathrm{Rm}~\Delta\Omega~\mathrm{rad} = \frac{\Delta \Omega_\mathrm{r} D^2}{\eta}
\end{equation}
were $R=0.85R_\odot$ is the radial location of the mid-convection
zone, $D=0.3R_\odot$ is the depth of the convection zone, $\Delta
\Omega_\mathrm{lat,r}$ are the angular velocity contrasts in latitude
and radius respectively \citep[e.g.,][]{Brown_et_al_2010a} 
and $\eta$ is the magnetic diffusivity at mid-convection zone.  These
magnetic Reynolds numbers are shown in Figure~\ref{fig:parameter
  space}$b$ for many of the dynamos, neglecting some cases with very
high magnetic Reynolds numbers.  

The latitudinal shear is generally large in all of these  dynamos
(horizontal axis), and all of the simulations, including those that
fail to sustain dynamo action, succeed in initially producing
global-scale toroidal magnetic structures.   The radial shear near the
equator is relatively weaker (vertical axis), and this quantity more
clearly separates those dynamos that succeed from those that fail.
The radial differential rotation also discriminates the cyclic dynamos
from those that build persistent fields.  Somewhat surprisingly, the magnetic Reynolds number
associated with the fluctuating convection does not provide as good of
a discriminant between dynamos that succeed or fail. The simulations
that fail to sustain dynamo action are those that do not regenerate
their poloidal fields quickly enough.  The clear dependence on
$\Delta\Omega_\mathrm{r}$ and the weak dependence on the properties of
the fluctuating convection suggest again that these wreath-building
dynamos may rely on effects other than a classical $\alpha$-effect to
build their turbulent emf $E_\mathrm{FI}$ which generates the
global-scale poloidal fields.

Two of the dynamos shown in Figure~\ref{fig:parameter space}$b$ have
negative magnetic Reynolds numbers.  These are the two slowly-spinning
simulations which rotate half as quickly as our Sun currently does
(e.g., case~D0.5).  In these simulations, the differential rotation is
anti-solar in nature and opposite in sense to that of the Sun, with
rapidly spinning pole and a more slowly spinning equator.  Despite
this fundamental difference, these simulations drive strong dynamos
and build magnetic wreath-like structures in their convection zones.
Anti-solar differential rotation appears to arise when convection is
only slightly constrained by rotation (e.g., when the Rossby number is
large), while solar-like differential rotation arises in rapidly
rotating stars (when the Rossby number is small).  The angular
velocity shear associated with the differential rotation increases as
the Rossby number becomes either large or small.  The Sun itself
appears to be very near Rossby number unity, and this partially
explains the difficulty in attaining wreath-building dynamos in
previous solar dynamo simulations: the angular velocity contrast in
the Sun is smaller than that realized in the rapidly rotating dynamos.
As a consequence, the solar dynamo simulations require low values of
$\eta$ to build wreaths, which in turn calls for high resolutions and
that exacts a large computational cost. 

\begin{figure}[!t]
\begin{center}
\plotone{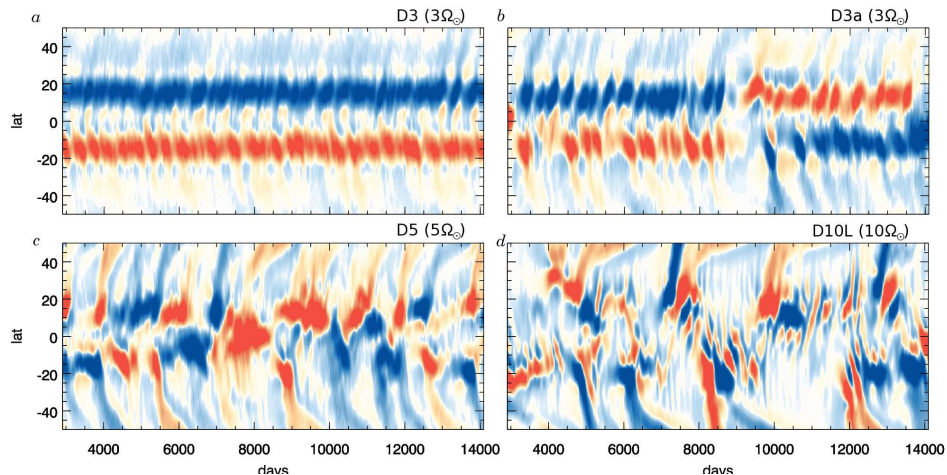}
\caption{Mean toroidal magnetic field in four wreath-building dynamos
 at mid-convection zone, shown as time latitude maps.  
 $(a)$~Persistent case~D3 with fields that do not change in sense. 
 $(b)$~More turbulent companion case~D3a shows long cycles. 
 $(c)$~Cyclic case~D5. $(d)$~Case~D10L.
}
 \label{fig:4 cases}
\end{center}
\end{figure}

Many of these dynamos show global-scale reversals, but the dependence
of this phenomena on rotation rate or Reynolds number are somewhat
unclear.   Near the onset of wreath-building dynamo action we
generally find little time variation in the axisymmetric magnetic
fields associated with the wreaths.  This is illustrated for case~D3 in
Figure~\ref{fig:4 cases}$a$, where the mean longitudinal field
$\langle B_\phi \rangle$ is shown at mid-convection zone over an
interval of nearly 10,000 days.  Though small variations are visible
on a roughly 500  day timescale, the two wreaths retain their
polarities for the entire time simulated (more than 20,000 days),
which is significantly longer than the convective overturn time
(roughly 10--30 days), the rotation period (9.3 days), or the ohmic
diffusion time (about 1300 days at mid-convection zone).  We refer to
the dynamos in this regime as persistent wreath-builders. 

Generally, we find that wreath-building dynamos begin to show large
time dependence as the magnetic diffusivity $\eta$ decreases and as
the rotation rate $\Omega$ increases.  Case~D3a, rotating three times
the solar rate but with lower diffusivities than case D3, is an example of the
first behavior and undergoes reversals even though D3 did not
(Fig.~\ref{fig:4 cases}$b$).  To explore the dependence on rotation
rate, we compare cases D3a, D5 and D10L, which have the same magnetic,
momentum and thermal diffusivities but rotate at three, five and ten
times the solar rate respectively.  In case D5 the global-scale reversals
are more frequent,  occurring with a roughly 1500 day timescale
(Fig.~\ref{fig:4 cases}$c$), though during some intervals the dynamo
can fall into other states.   When we explore case D10L rotating ten
times the solar rate, we find that the cycles are somewhat harder to
define, with the northern and southern hemispheres showing distinctly
different behavior (Fig.~\ref{fig:4 cases}$d$).  It is unclear at
present how the cycle period depends on the rotation rate of the star;
case D3a and D5 would imply that faster rotation leads to shorter
cycles in general agreement with observations, but all of these
simulations are highly variable and actual cycle periods are difficult
to quantitatively define.

The magnetic wreaths act back strongly on the differential rotation
that feeds their generation, and the global-scale shear is much weaker
in dynamo simulations than in corresponding hydrodynamic simulations.
Individual convective structures are 
largely unaffected by the magnetic wreaths except when the fields
reach very large amplitudes; in case D5 this occurs when $B_\phi$
exceeds values of roughly 35~kG at mid-convection zone.  At the
highest rotation rates the Lorentz force of the axisymmetric magnetic
fields becomes strong enough to substantially modify the differential
rotation, largely wiping out the latitudinal and radial shear (e.g.,
cases D10 and D15 in Fig.~\ref{fig:parameter space}$a$).  In these
cases, wreath-like structures can still form though they typically
have more complex structure and are less axisymmetric.

\section{Overview}
\label{sec:challenges}

Advances in massively parallel supercomputers are now permitting
simulations that can capture global-scale convection and dynamo action
in stars like our Sun.  Dynamo simulations of solar-type stars are
revealing that organized magnetic fields can be built in the
convection zone itself, without necessarily relying on a tachocline
in between the convection zone and radiative zone for this
organization.  This is a marked departure from many solar dynamo
theories, where the tachocline plays a vital role. 
Many of these simulations show quasi-cyclic reversals
of magnetic polarity.  These cycles are not yet like the solar cycle:
namely, they are typically too short and generally the magnetic fields
migrate towards the poles, rather than towards the equator as observed
at the solar surface \citep[though see][]{Ghizaru_et_al_2010}.
Simulations remain well separated in parameter space from real stellar
convection, which remains humblingly out of reach for the foreseeable
future, but these global-scale simulations are entering a regime
where resolved turbulence plays a larger role than explicit
diffusion.  Thus they are beginning to capture in a self-consistent
fashion the processes which likely contribute most directly to stellar
dynamo action.

In the future we will be exploring convection and dynamo action in K-
and F-type stars, to understand how stellar mass and convection zone
depth affect the global-scale dynamo.  This work will complement
ongoing work exploring dynamo action in fully convective M-dwarfs
\citep{Browning_2008}.  As simulations move away from the Sun, we need
further constraints from stellar observations.  In particular,
measurements of stellar differential rotation are vitally important,
given the role of that global-scale flow in the wreath building
dynamos.  An observational understanding of how dynamo properties
including magnetic activity and cyclic period scale with differential
rotation would be of great utility.  
  
\acknowledgements
We thank Nicholas Nelson for his excellent and continuing work on dynamo
case D3a (Figure~\ref{fig:4 cases}$b$) and its relatives.
This research is supported by NASA through Heliophysics Theory
Program grants NNG05G124G and NNX08AI57G, with additional support for
Brown through the NSF Astronomy and Astrophysics postdoctoral fellowship
AST 09-02004. CMSO is supported by NSF grant PHY 08-21899.
Miesch was supported by NASA SR\&T grant NNH09AK14I.
NCAR is sponsored by the National Science Foundation.
Browning is supported by research support at
CITA.  Brun was partly supported by the Programme National Soleil-Terre
of CNRS/INSU (France), and by the STARS2 grant from the
European Research Council. The simulations were carried out with NSF
PACI support of PSC, SDSC, TACC and NCSA.   Field line tracings shown
in Figure~\ref{fig:case D5} were produced using VAPOR  \citep{Clyne_et_al_2007}.

\bibliography{bibliography}

\end{document}